**Perceived community alignment increases information sharing**

Elisa C. Baek[1,2]*, Ryan Hyon[2], Karina López[2], Mason A. Porter[3,4,5], and Carolyn Parkinson[2,6]*

[1]Department of Psychology, University of Southern California, Los Angeles, CA, United States of America, [2]Department of Psychology, University of California, Los Angeles, Los Angeles, CA, United States of America, [3]Department of Mathematics, University of California, Los Angeles, Los Angeles, CA, United States of America, [4]Department of Sociology, University of California, Los Angeles, Los Angeles, CA, United States of America, [5]Sante Fe Institute, Santa Fe, NM, United States of America, [6]Brain Research Institute, University of California, Los Angeles, Los Angeles, CA, United States of America

* Corresponding authors:

Elisa C. Baek: elisa.baek@usc.edu

Carolyn Parkinson: cparkinson@ucla.edu

## Abstract

It has been proposed that information sharing, which is a ubiquitous and consequential behavior, plays a critical role in cultivating and maintaining a sense of shared reality. Across three studies, we tested this theory by investigating whether or not people are especially likely to share information that they believe will be interpreted similarly by others in their social circles. Using neuroimaging while members of the same community viewed brief film clips, we found that more similar neural responding of participants was associated with a greater likelihood to share content. We then tested this relationship using two behavioral studies and found (1) that people were particularly likely to share content that they believed others in their social circles would interpret similarly and (2) that perceived similarity with others leads to increased sharing likelihood. In concert, our findings support the idea that people are driven to share information to create and reinforce shared understanding, which is critical to social connection.

## Introduction

Information sharing is a ubiquitous human behavior. Interpersonal sharing of information, which can spread particularly effectively in online media, can powerfully shape people's opinions, behaviors, and attitudes across domains (ranging from health behaviors[1] to political action[2]). Additionally, it has been hypothesized that information sharing supports fundamental human motivations to connect and belong socially[3,4] and also plays an important role in constructing and reinforcing a sense of generalized shared reality (i.e., the sense of "being on the same page"), which is critical for social connection[5,6].

Corroborating the above two hypotheses, empirical evidence suggests that anticipation of positive social interactions is a key motivation for sharing information[7,8] and recent neuroimaging work has demonstrated that activity in regions of the brain that are involved in mentalizing (i.e., understanding the mental states of others) plays an important role in information sharing. For example, regions of the brain that are associated with mentalizing (e.g., the medial prefrontal cortex, precuneus, temporal junction, and superior temporal sulcus[9,10]) are activated when people think about sharing content with others[11]. Accordingly, when people make sharing decisions, they may spontaneously consider how others would respond to the shared information. The extent to which a piece of content engages these brain regions is associated both with neuroimaging participants' self-reported likelihoods of sharing[11] and with population-level virality (i.e., how often the content is actually shared in the real world)[12]. Behavioral evidence also suggests that the relationship between mentalizing and sharing likelihood is causal; thinking about other people's mental states and perspectives when considering content to share increases the likelihood of sharing content[13]. Collectively, these results suggest that people actively consider the mental states of other people when considering content to share and are motivated to share information to fulfill their needs to connect socially with others.

Given that having shared understanding with others is linked to social connection[14,15] and that the desire to connect socially is a key motivation for sharing behavior[7,8], one possibility is that people consider the extent to which content will cultivate shared understanding with others when deciding whether or not to share it. Therefore, the involvement of mentalizing processes in information sharing may, in part, reflect individuals considering the perspectives of potential information receivers to determine whether or not others would respond to the shared information in ways that evoke shared understanding. For instance, people may share information that they believe others will interpret similarly because doing so reinforces perspectives, attitudes, and beliefs about the world that are already well-established and agreed upon in their social circles; shared understanding across these various facets is important to social connection[4,14,15].

In the present paper, we investigate the idea that motivations to achieve and maintain shared reality with others may play a critical role in information sharing. We thereby provide empirical evidence that advances existing theories about the motivations behind information sharing, which have often focused on non-social drivers of sharing (e.g., the desire to spread information that fulfills a need for accuracy[16–20]). Across three studies, we test the hypothesis that people are more likely to share information when they believe that others in their social circles will share their viewpoints and opinions about the information than when they believe that others' viewpoints will differ from theirs.

In Study 1, we used functional magnetic resonance imaging (fMRI) to test whether or not people are more likely to share content when it evokes similar neural responses in members of their social circles. We used inter-subject correlations (ISCs) of neural responses while participants watched dynamic, naturalistic stimuli (i.e., videos) to capture the similarity of brain responses across participants as these responses unfold over time. Prior research has linked ISCs of neural responses to naturalistic messages with participants' interpretations and understanding of messages[21–23], suggesting that this approach can meaningfully capture similarities in relevant high-level psychological processes (e.g., inferring others' mental states or integrating incoming information into existing knowledge) across individuals.

The results of Study 1 support our hypothesis. We found that coordinated neural responses in brain regions that previously have been implicated in shared high-level interpretations and low-level sensory processing are associated with an increased sharing likelihood, suggesting that similarities in interpretations and understanding of messages across individuals are associated with the likelihood of sharing the messages. Building from the results of Study 1, we directly tested these associations by examining whether or not individuals are more likely to share content when they believe that others in their social circles will interpret the content similarly to themselves. Accordingly, we conducted an online behavioral study (Study 2) and found that participants were especially likely to share content when they believed that other people in their social circles would have similar views about the content as themselves. These results held even when controlling for participants' levels of interest in the content and for their evaluations of it. We then conducted an experimental study (Study 3) to test whether or not a general sense of similarity with others causally increases the sharing likelihood. The results of Study 3 suggest that this relationship is causal, with perceived alignment between one's broad attitudes and preferences and those of others in one's social circles causally increasing the sharing likelihood.

Taken together, the findings of our three studies suggest that people are more likely to share information when they believe that others in their social circles share their own viewpoints and opinions.

## Results

### Study 1: fMRI study

**Neural similarity.** During the fMRI study, participants watched a set of video clips on a variety of topics. For details, see the Methods section and Supplementary Table 1. All participants were living in one of two social communities of a first-year dormitory in a large public university in the United States. This allowed us to test whether or not people are especially likely to share content that members of their own community interpret similarly, as indicated by their neural responses. In each brain region (see the Methods section for details about the parcellation and the preprocessing of the fMRI data), we computed the Pearson correlation between the time series of neural responses for each pair of participants (i.e., dyad) for each video. This yields one correlation coefficient for each unique combination of dyad, video, and brain region. See the Methods section for more details.

**Sharing-likelihood ratings.** After the fMRI portion of Study 1, participants indicated their likelihood of sharing each video on social media on a 1–5 Likert scale (with "1 = very unlikely" and "5 = very likely"). In our primary analyses, we binarized the sharing-likelihood ratings (see the Methods section). This choice is consistent with recent studies that link neural similarity with behavioral measures[24–26]. To relate the participant-level sharing-likelihood ratings with the dyad-level neural-similarity measure, for each video, we transformed the participant-level binarized sharing-likelihood measure into a dyad-level sharing-likelihood measure. For each video, we categorized a dyad's sharing-likelihood rating as (1) {high sharing, high sharing} if both participants in the dyad had a high likelihood of sharing the video, (2) {low sharing, low sharing} if both participants in the dyad had a low likelihood of sharing the video, and (3) {low sharing, high sharing} if one participant had a high likelihood of sharing the video and the other had a low likelihood of sharing it. Unlike existing studies, which have investigated whether or not similarities in a participant-level attribute (e.g., their number of friends[21] or loneliness[27]) are linked with greater neural similarity, we are interested in whether or not people are more likely to share content when it evokes similar neural responses in individuals in their social circles. Therefore, our contrasts are at the level of video–dyad combinations.

For each brain region, we fit a linear mixed-effects model with crossed random effects to account for the dependence structure of the data[28] (see the Methods section) with the ISC in the corresponding region as the dependent variable, the dyad-level sharing-likelihood measure as the independent variable, and similarities in participants' age, gender, and country of origin as control variables. See Supplementary methods 1 for more details. We then conducted a planned-contrast analysis[29] to identify brain regions for which a high sharing likelihood is associated with more coordinated neural responses than a low sharing likelihood (i.e., $\text{ISC}_{\{\text{high sharing, high sharing}\}} > \text{ISC}_{\{\text{low sharing, low sharing}\}}$). We focus on the contrast $\text{ISC}_{\{\text{high sharing, high sharing}\}} > \text{ISC}_{\{\text{low sharing, low sharing}\}}$, as this contrast is our most direct test of the hypothesis that people are more likely to share content that different individuals interpret similarly than to share content that different individuals do not interpret similarly. In Supplementary Fig. 1, we show our results for our exploratory contrasts $\text{ISC}_{\{\text{high sharing, high sharing}\}} > \text{ISC}_{\{\text{low sharing, high sharing}\}}$ and $\text{ISC}_{\{\text{low sharing, high sharing}\}} > \text{ISC}_{\{\text{low sharing, low sharing}\}}$. We employed Holm–Bonferroni correction to correct for multiple comparisons across brain regions for each contrast. We also performed analyses to examine the relationships between a non-binarized version of the sharing-likelihood ratings and neural similarity.

**Results of Study 1.** There were larger ISCs in the temporoparietal junction, superior parietal cortex, and regions of the visual cortex when participants were very likely to share information (i.e., {high sharing, high sharing}) than when participants were unlikely to share information (i.e., {low sharing, low sharing}) (see Fig. 1c). (In Supplementary Table 2, we give our complete set of results for subcortical brain areas.) We observed a similar pattern of results in our exploratory contrasts (i.e., $\text{ISC}_{\{\text{high sharing, high sharing}\}} > \text{ISC}_{\{\text{low sharing, high sharing}\}}$ and $\text{ISC}_{\{\text{low sharing, high sharing}\}} > \text{ISC}_{\{\text{low sharing, low sharing}\}}$; see Supplementary Fig. 1), for our analyses with a non-binarized version of the sharing-likelihood variable (see Supplementary Fig. 2), and for our analyses using an alternative statistical-modeling approach (see Supplementary Fig. 3).

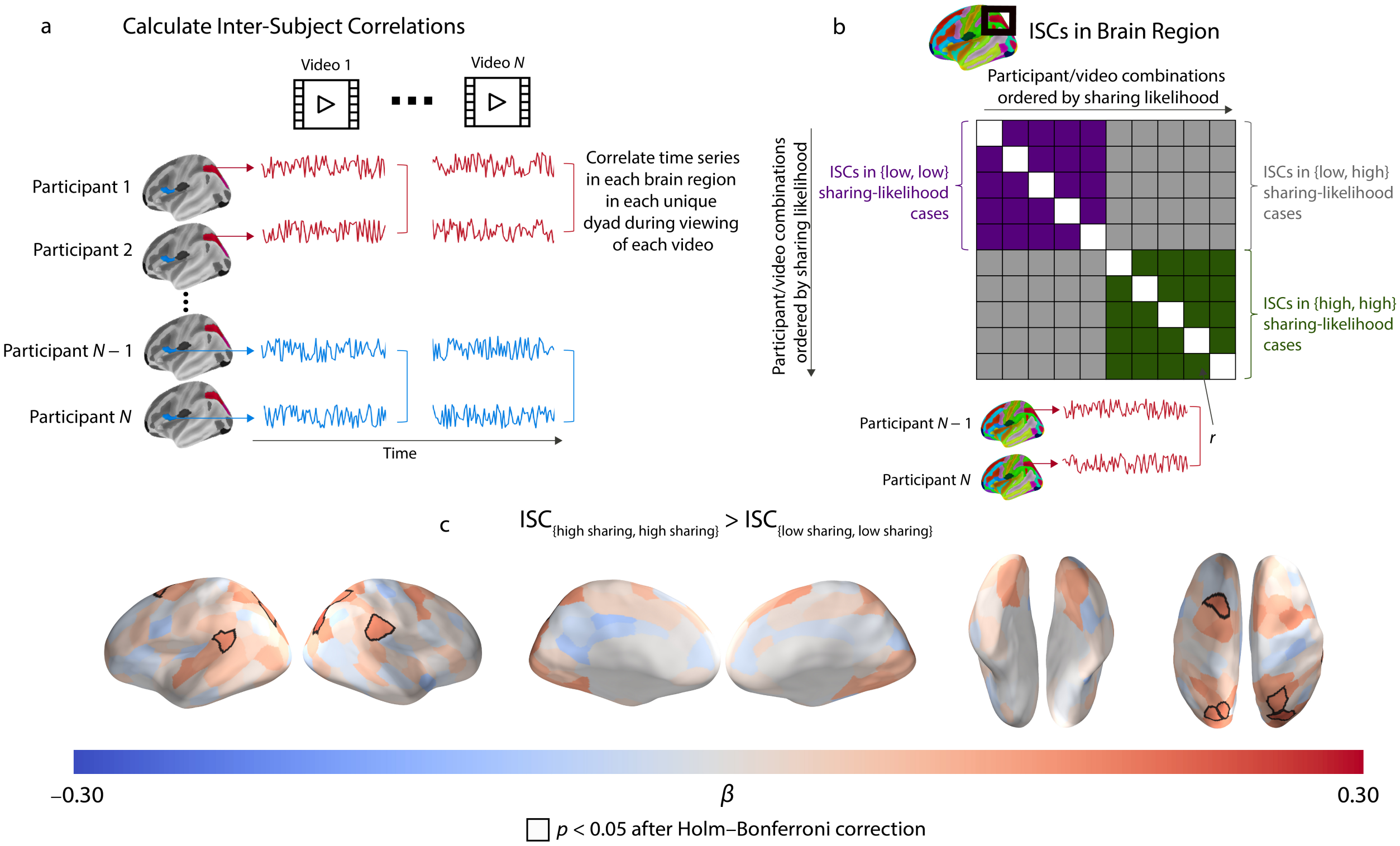

**Fig 1. Similar neural responses in members of a social community are associated with increased likelihood of information sharing. (a)** We extracted time series of neural responses while participants watched each video. For each unique dyad (i.e., pair of participants), we calculated inter-subject correlations (ISCs) of these time series for each of the 214 brain regions for each video. **(b)** We related neural similarity with participants' self-reported

likelihood of sharing the videos. Each cell of the matrix consists of the ISC between two participants for a brain region. The rows and columns of the matrix are ordered by participants' sharing-likelihood ratings. We performed planned contrasts of the different sharing-likelihood ratings to test whether or not there was a larger ISC when both individuals in a dyad indicated a high sharing likelihood (i.e., $ISC_{\{\text{high sharing, high sharing}\}}$) than when both individuals in a dyad indicated a low sharing likelihood (i.e., $ISC_{\{\text{low sharing, low sharing}\}}$). **(c)** There were larger ISCs in the temporoparietal junction, superior parietal cortex, and regions of the visual cortex when participants were very likely to share than when participants were unlikely to share. The quantity $\beta$ is the standardized regression coefficient. [The figures in panels (a) and (b) are adapted from prior work[24,26].]

## Study 2: Correlational behavioral study

The results of Study 1 demonstrate that similar neural responses of individuals in a social community are associated with a greater likelihood of sharing content. Combined with previous observations that decisions to share content involve the brain's mentalizing system[11,12], these results are consistent with the possibility that people may be driven to share content when they believe that others in their social circles will similarly interpret that content. Notably, the results in Study 1 have potential alternative explanations. For example, when an individual finds that particular content is engaging, there can be both less mind-wandering (and hence greater alignment with others' neural responses[30]) and a greater desire to share that content. This latter possibility does not require participants to be aware that the content that they rate as more worthy of sharing also elicits similar responses in others. When content is engaging, people may both have especially similar neural responses to it and be particularly likely to share it with others without necessarily realizing that the content may evoke very similar responses across perceivers. Therefore, in Study 2, we directly tested the hypothesis that people are more inclined to share content to which they believe that others in their social circles will have similar responses through a pre-registered online behavioral study of 100 participants. (The preregistration is at https://osf.io/qm4zw.) In this study, participants rated news articles on the extent to which they believed others in their social circles would share their views about the content (on a scale with the anchors "these people may or may not share my view" and "I am confident that most of these people would share my view"), how likely they were to share each article on social media, the extent to which they believed that their social-media friends would find the article interesting, and the extent to which they believed that their social-media friends would find the article positive or negative (i.e., its valence). To address limitations in Study 1 from the fixed order of the stimuli and the time gaps between stimulus presentations and sharing-likelihood ratings, we randomized the order of the stimuli in Study 2. Additionally, participants answered questions about their likelihood to share each piece of content shortly after viewing the stimuli. See the Methods section for more details.

**Results of Study 2.** To test our hypothesis that people are more likely to share content that they believe will be interpreted similarly by others in their social community, we fit a linear mixed-effects model to account for the dependence structure of the data (see the Methods section) with sharing likelihood as the dependent variable and perceived-similarity rating as the independent variable. We found a positive association between perceived similarity and sharing likelihood ($\beta = 0.398$; SE = 0.041, $p < 0.001$; see Fig. 2), indicating that participants were more likely to share information when they believed that others in their social circles would share their views about the content. Given prior work that suggests links between information sharing and both the valence of content and the extent to which content is perceived as interesting[17,31,32], we also fit a linear mixed-effects model with sharing likelihood as the dependent variable, perceived-similarity rating as the independent variable, and participants' interest and valence ratings as control variables. We found that the association between perceived similarity and sharing likelihood remained significant even after controlling for interest and valence ratings ($\beta = 0.189$, SE = 0.038, $p < 0.001$). This suggests that the link between perceived similarity and sharing likelihood does not arise merely because people are more likely to share and to have similar perceptions of information that is more interesting, extremely positive, or extremely negative. The results of Study 2 support our interpretations of our neuroimaging findings from Study 1, suggesting that people are more likely to share content that they believe will evoke similar interpretations across different individuals.

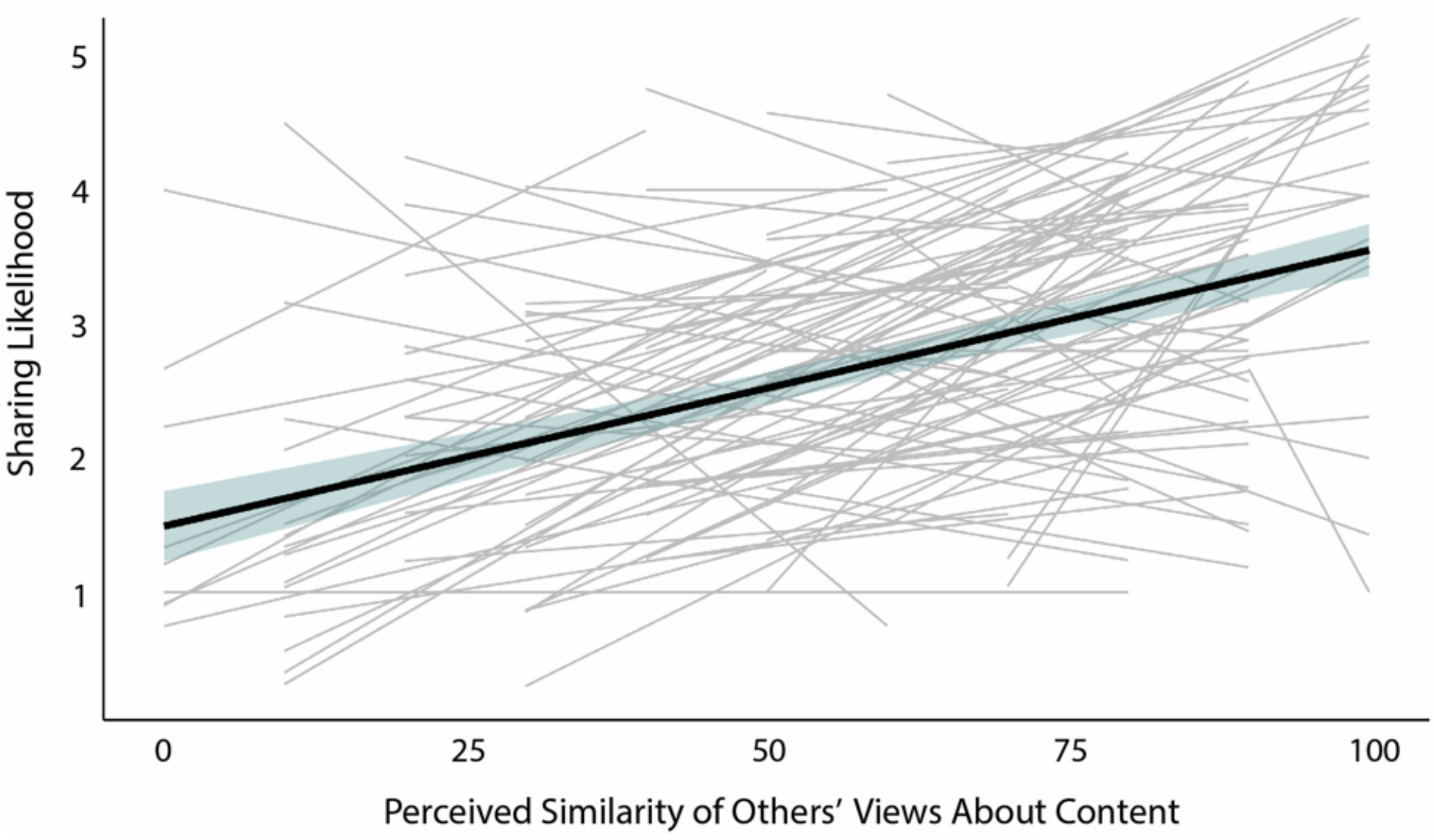

**Fig 2. Participants are more likely to share content when they believe that others will interpret the content similarly as themselves.** There was a positive association between perceived similarity and sharing likelihood in Study 2. That is, the study participants were more likely to share information when they believed that others in their social circles would share similar views of the content as themselves. The black line gives the mean group-regression line, the light blue band indicates the 95% confidence interval, and the light gray lines are participant-level regression lines. We measured perceived similarity on a scale with the anchors "0 = these people may or may not share my view" and "100 = I am confident that most of these people would share my view." See the Methods section for more details.

## Study 3: Behavioral experiment

Given the results of Study 2, which suggest that there is an association between the perceived similarity of others' views about a specific piece of content and their likelihood to share that content, we tested whether or not a general sense of similarity with others causally increases the sharing likelihood. Accordingly, we conducted an online experimental study (Study

3) to test whether or not participants are more likely to share information on social media with others who broadly hold similar views and preferences as themselves than with others who hold dissimilar views and preferences. This builds on Study 2 to test the theory that participants are more likely to share information with others who tend to share similar beliefs, preferences, and traits as themselves because presumably such similarly-minded people are also more likely to share their views on diverse types of content. (The preregistration is available at https://osf.io/7tvcb.) It is possible that asking participants about sharing likelihood and perceived similarity in close succession in Study 2 increased the chance that participants gave similar responses to both questions. Study 3 alleviates this concern by experimentally manipulating perceived similarity and having participants report only their sharing likelihoods.

In this study, 300 participants first answered a series of questions about their demographic information and their preferences about a variety of topics (e.g., movies, news sources, and television shows). (See the Methods section for more details.) Participants were then given a choice of five news articles and selected the article in which they were most interested. They were then assigned uniformly at random to one of four experimental conditions. In each condition, participants were asked to consider sharing the news article with a Facebook group with a different social context[33]: (1) participants in the "similar social context" condition were told that the majority of other people in the Facebook group were similar to them in demographic traits and preferences; (2) participants in the "dissimilar social context" condition were told that the majority of other people in the group were dissimilar to them; (3) participants in the "unclear social context" condition were told that it was not clear whether or not other people in the group shared their demographic traits or preferences; and (4) participants in the "mixed social context" condition were told that some people in the group were similar to them and others were different from them in their demographic traits and preferences. All participants were then asked to indicate their likelihood of sharing the article that they had chosen earlier with the Facebook group.

Our main hypothesis in Study 3 was that participants would be more likely to share information with others who they believed were similar to themselves in views, preferences, and demographic traits (and hence presumably would respond similarly to content) than with others who they believed were different from themselves in views, preferences, and demographic traits. To test this hypothesis, we first fit a linear regression model with sharing likelihood as the dependent variable and the experimental condition (i.e., social context) as the independent variable. We then performed a planned-contrast analysis[29] to test whether or not there was a greater sharing likelihood when participants considered sharing the content with others who they believed had very similar views, preferences, and demographic traits to their own than when they considered sharing with others who they believed were dissimilar to themselves (i.e., similar > dissimilar). Given that individual differences in baseline sharing (i.e., how often an individual generally shares content online) and level of interest in the content are likely to affect participants' sharing likelihood, we also fit an additional model and performed a planned-contrast analysis with baseline sharing and interesting ratings as control variables. Furthermore, although the similar > dissimilar contrast is the most direct test of our main hypothesis, we explored whether or not participants would be more likely to share information with a group of similar others than a group of others who they believed had mixed traits, views, and preferences (i.e., similar > mixed) or a group of others in which it was unclear whether or not they shared their traits, views, and preferences (i.e., similar > unclear). We report the results of all possible contrasts in Supplementary Tables 3 and 4. For all of our analyses, we employed false-discovery-rate (FDR) correction to correct for multiple comparisons due to multiple contrasts.

**Results of Study 3.** As hypothesized, we found that participants were more likely to share information with others who they perceived as similar than with others who they perceived as dissimilar (i.e., similar > dissimilar) ($\beta = 0.572$, SE = 0.158, $p_{corrected} = 0.001$; see Fig. 3). The results held even when controlling for participants' baseline sharing and interesting ratings ($\beta = 0.854$, SE = 0.192, $p_{corrected} < 0.001$). We also found that participants were more likely to share information with a group of others who they perceived as similar than with a group of others in which they perceived some people as sharing their views and others as not sharing them (i.e., similar > mixed) ($\beta = 0.289$, SE = 0.158, $p_{corrected} = 0.092$), although this relationship is only marginally statistically significant. Participants were also more likely to share information with a group of others who they perceived as similar than with a group in which it was unclear whether or not the people in it shared their views (i.e., similar > unclear) ($\beta = 0.583$, SE = 0.159, $p_{corrected} = 0.001$).

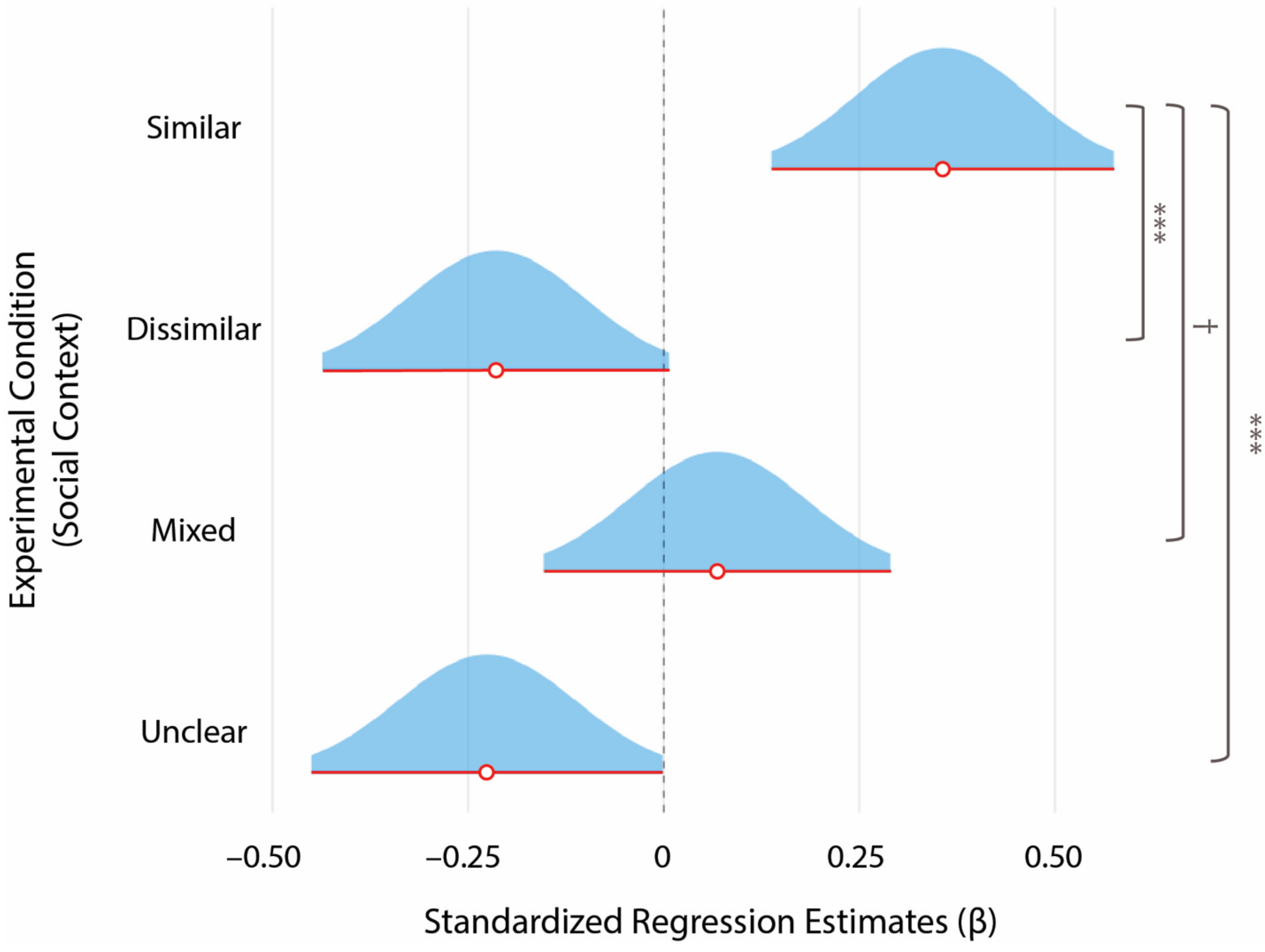

**Fig 3. Perceived community alignment increases participants' sharing likelihood.** Participants in Study 3 were more likely to share information with a group when they believed that the people in that group had similar demographic traits, views, and preferences as themselves and presumably would respond similarly to content as themselves (i.e., similar > dissimilar). Participants were also more likely to share with a group of others who they perceived as similar than with a group in which it was unclear whether or not the people in it were similar to themselves (i.e., similar > unclear). Participants were also more likely to share with a group of others who they perceived as similar than with a group in which they perceived some people as similar and others as dissimilar (i.e., similar > mixed), although this difference is only marginally statistically significant. See Supplementary Tables 3 and 4 for the results of all examined contrasts. The white circles indicate regression estimates from a linear model that predicts sharing-likelihood ratings from experimental condition (i.e., social context). The red lines indicate 95% confidence intervals of the estimates, and the blue regions indicate the associated distributions. The symbol *** denotes a p-value of $p < 0.001$, and the symbol † denotes a p-value of $p < 0.01$

## Discussion

What drives information sharing? Across three studies, we found that people are more likely to share information when they believe that the information will be interpreted similarly by others in their social community. We found that inter-subject neural similarity in several regions of the brain, including regions that are associated with both low-level sensory and high-level cognitive processing, was correlated with sharing likelihood. Accordingly, our findings suggest that information is more likely to be shared when it engages individuals' brains in similar ways, capturing their attention in a coordinated fashion. In concert with prior work that highlights the involvement of the brain's mentalizing system during decisions to share information[11,12,34], our results suggest that people's decisions to share content may be driven partly by the extent to which they believe that it will be processed similarly by others in their social communities. Indeed, our behavioral studies that directly test these relationships give strong evidence that perceived similarity causally increases information sharing. Specifically, we found that people were more likely to share content when they believed that others would share their viewpoints and opinions about it. Taken together, our findings are consistent with theories of information sharing as an inherently social behavior that plays a critical role in forming and reinforcing shared realities, which in turns promotes social connection and cohesion[3,5].

Brain areas in which coordinated activity was associated with increased sharing likelihood included regions of the temporoparietal junction that are part of the default mode network. These regions have been implicated previously in social cognitive processes such as mentalizing (e.g., taking the perspective of others)[9,10,35], and the magnitude of the brain activity in these regions has been linked to both individual and population-level sharing behavior of short text-based content[11,12]. Our work extends these findings to show that the extent to which complex, dynamic messages evoke greater coordinated activity in these regions is linked to the likelihood that content is shared. Furthermore, as inter-subject similarity of neural responses in regions of the default mode network has been associated with shared interpretations and understanding of narratives[21,22], one interpretation of our results is that people are more likely to share content that evokes a sense of collective meaning in their social environment. Accordingly, our results align with prior work that found that people are more likely to share content that they believe will strengthen their social relationships[17,36]. Our work also suggests that one way that people do so is by sharing content that reinforces agreed-upon attitudes and beliefs.

We also found an association between greater sharing likelihood and inter-subject similarity of responses in brain regions that are associated with attention allocation (e.g., the superior parietal lobule) and low-level sensory cortices (e.g., regions of the visual cortex). One possibility is that messages that people feel are worthy of sharing capture and coordinate individuals' attentional processes. Indeed, there is evidence that neural responses in the dorsal attention network and low-level sensory cortices not only align when people are exposed to the same naturalistic stimuli[37,38], but also coordinate across individuals to the extent that they exhibit similar higher-level processing of the stimuli[39,40]. Accordingly, our findings that implicate similarities in the brain's higher-level cortical systems, such as regions that are involved in attention allocation and regions of the default mode network, in increased sharing likelihood suggest that this alignment of the low-level sensory regions may be due to top-down modulations that are driven by attentional and social motivations[41–44].

The results of Study 1 results suggest that similar neural responses across individuals in a social community is associated with a greater sharing likelihood. In conjunction with prior observations that decisions to share information involve the brain's mentalizing system[11,12,34], these results suggest that people may be driven to share content when they believe that others in their social circles will interpret and respond to the content similarly to themselves. In two pre-registered follow-up studies (Studies 2 and 3), we directly tested whether or not individuals are more likely to share content when they believe that others in their social circles will interpret the content similarly to themselves. We thereby directly tested our hypothesis against potential alternative explanations of the neural results (for instance, that content that is more vivid or exciting may entrain brain activity and also be more likely to be shared, regardless of whether or not participants believed that others would view the content similarly). In Study 2, we found that people were more likely to share content when they believed that others in their social circles would have similar views as themselves about it. In Study 3, we found evidence that a general sense of similarity with others causally increases the sharing likelihood, presumably in part because such similarly-minded others may also be more inclined to share their viewpoints on a variety of topics. Specifically, we found that people were more likely to share content when they perceived that potential receivers of that content held similar views as themselves than when they perceived that the potential receivers held dissimilar or unclear views. Accordingly, the results from our three studies corroborate theories of information sharing as an inherently social behavior[5,13,45] that supports fundamental human motivations to connect and belong[3], rather than theories that emphasize non-social motivations (such as a desire for accuracy)[16–20]. Given that shared understanding is important to social connection[14,15,26], our findings suggest that, by sharing information, individuals create and establish collective meaning that promotes social connection through shared worldviews with others around them.

The stimuli in our studies included a variety of different topics and themes (e.g., a scientific demonstration, comedy clips, and social issues in Study 1; see Supplementary Table 1). Therefore, we are unable to make strong claims about specific message-level characteristics that may influence the effects that we found between perceived similarity and sharing likelihood. However, our results illustrate that content—regardless of its specific theme or domain—is more likely to be shared when individuals expect others to interpret the content similarly to themselves. We see this in the coordinated neural responses in Study 1, the self-report data in Study 2, and the experimental manipulation in Study 3. Our findings highlight fundamental neurobiological and psychological processes that motivate and predict sharing behavior across different content characteristics. Future work that explores these effects for different types of content (e.g., political content, morally-charged content, controversial content, and others) can further test whether the relationship between perceived similarity and sharing likelihood is affected by the content type (e.g., if these effects are heightened or reduced in certain contexts).

In Study 1, the videos were not presented in isolation; instead, they were presented in a fixed order amidst a stream of other content. Therefore, comparisons of the 14 videos in Study 1 may have influenced participants' likelihood to share. Although this setting has analogues in daily life experiences, where individuals watch videos on a variety of social media (e.g., TikTok, YouTube, and Instagram) in sequences that are affected by platforms' algorithms and still compare pieces of content to one another when determining shareworthiness, future work can help elucidate the effects of contextual factors (such as the order in which stimuli are presented) on sharing likelihood.

It also remains unclear whether our results still apply in contexts in which individuals have overt motivations to seek different viewpoints from their own when sharing content (e.g., when one seeks critiques of content or is unsure of how to interpret content). In such contexts, perceived similarity in viewpoints with others may not be a key driver of information sharing. Furthermore, Study 1 participants were young adults, and Studies 2 and 3 used online convenience samples in the United States. Future work can clarify whether our findings generalize across diverse contexts and populations. Moreover, future studies that include other forms of information sharing that do not involve social media may provide further insight into whether our findings also hold for other sharing contexts (e.g., offline sharing of information by people who are not regular users of social media).

Our findings also have potential applications for studying various consequential phenomena in information sharing. For instance, one can use the links between perceived similarity and sharing likelihood as a theoretical framework to study the motivations that lead to the spread of misinformation, which has widespread negative consequences[46,47]. One potential future direction is testing whether individuals' proclivity to share information when it evokes similar responses across members of their social circles may cause them to be insufficiently concerned about the accuracy of content before sharing it. One can also use a theoretical framework that is based on our results to improve the design of messaging campaigns. For instance, public-service announcements that are more likely to be interpreted similarly across individuals in a social community may be more likely to lead to message-congruent behavior, which can have positive impact for pro-social and pro-health messages. Indeed, in one study, similarity in neural responses in a small group of participants was associated with real-world engagement levels of media content[48]. Additionally, effective speeches elicit more similar neural responding across individuals than ineffective speeches[49]. It seems particularly fruitful for future work to explicitly test whether or not similarly-interpreted content is more effective and more likely to be shared.

In summary, our results suggest that individuals are more likely to share information when they believe that it will be interpreted similarly by others in their social circles. We found that coordinated neural responses across individuals was associated with increased sharing. In subsequent behavioral studies, we found convergent evidence that individuals were more likely to share content when they believed that others in their social circles would hold similar viewpoints as themselves. In concert, our findings support the idea that information sharing plays a critical role in creating and reinforcing individuals' shared realities,-which is important to social connection.

## Methods

### Study 1: fMRI study

**Study participants.** A total of 70 participants participated in our fMRI study. All participants were living in one of two communities of a first-year dormitory in a large public university in the United States. We tested whether or not participants were more likely to share content that they felt would be interpreted similarly, as indicated by similar neural responses, by others in their social community. We excluded all data from four participants. One participant did not complete the scan, two participants had excessive head movement, and one participant fell asleep in the scan. Additionally, we included only partial data from two participants. One participant had excessive head movement in one of the runs, and one participant reported falling asleep in one of the runs. Therefore, of the 66 individuals in our analysis, we used full data from 64 of them and partial data from 2 of them. All participants provided informed consent in accordance with the procedures of the Institutional Review Board of the University of California, Los Angeles. We reported on separate analyses of the Study 1 data set in manuscripts that examined other (and very different) research questions[26,27,50].

**fMRI procedure**. Participants attended a study appointment that included a 90-minute session in which they were scanned using blood-oxygen-level-dependent (BOLD) fMRI and completed a series of self-report surveys. Prior to the fMRI portion of the study, participants completed a demographic survey, from which we obtained their self-reported ages and genders. We then informed participants that they would be watching a series of video clips in the fMRI scanner while their brain activity was measured. We also informed them that their experience would be akin to watching television while another person "channel-surfed"[i]. We instructed the participants to watch the videos naturally, as they would in real life. In the scanner, participants watched 14 video clips with sound that ranged in duration (from 91 to 734 seconds) and content. (See Supplementary Table 1 for descriptions of the content.) The video task was divided into four runs, and the total task lasted approximately 60 minutes. All participants saw the videos in the same order[ii]. After the fMRI scan, participants indicated their likelihood to share each video on social media by answering the question "How likely would you be to share this video on social media?" with the anchors "1 = very unlikely" and "5 = very likely" (as used in prior work[11]).

**fMRI data acquisition**. We acquired neuroimaging data using a 3T Siemens Prisma scanner with a 32-channel coil. The functional images were recorded using echo-planar sequences (with echo time = 37 ms, repetition time (TR) = 800 ms, slice thickness = 2.0 mm, voxel size = 2.0 mm × 2.0 mm × 2.0 mm, matrix size = 104 × 104 mm, field of view = 208 mm, multi-band acceleration factor = 8, and 72 interleaved slices with no gap between them). To allow stabilization of the BOLD signal, we added a "start" buffer (with a duration of 8 seconds) and an "end" buffer (of 20 seconds) to the beginning and end of each run, respectively. Participants saw a blank black screen during these buffers. We also acquired high-resolution T1-weighted (T1w) images (with echo time = 2.48 ms, repetition time = 1,900 ms, slice thickness =

---

[i] The term "channel-surfing" is an idiom that refers to scanning through different television channels.

[ii] We performed permutation tests and found that there was no significant relationship between sharing likelihood and when a video clip appeared in the stimulus sequence. See the Supplementary Material for more information.

1.0 mm, voxel size = 1.0 mm × 1.0 mm × 1.00 mm, matrix size = 256 × 256 mm, field of view = 256 mm, and 208 interleaved slices with a 0.5 mm gap between them) to use in coregistration and normalization. To minimize head motion, we attached adhesive tape to the headcase and stretched it across participants' foreheads[51].

**fMRI data analysis**. We used fMRIPrep version 1.4.0 for the data processing of our fMRI data[52]. We have taken the descriptions of anatomical and functional data preprocessing that begins in the next paragraph from the recommended boilerplate text that is generated by fMRIPrep and released under a CC0 license, with the intention that researchers reuse the text to facilitate clear and consistent descriptions of preprocessing steps, thereby enhancing the reproducibility of studies.

For each participant, the T1-weighted (T1w) image was corrected for intensity non-uniformity (INU) with N4BiasFieldCorrection, distributed with ANTs 2.1.0[53], and used as a T1w-reference throughout the workflow. Brain-tissue segmentation of cerebrospinal fluid (CSF), white matter (WM), and gray matter (GM) was performed on the brain-extracted T1w using FSL fast[54]. Volume-based spatial normalization to the ICBM 152 Nonlinear Asymmetrical template version 2009c (MNI152NLin2009cAsym) was performed through nonlinear registration with antsRegistration (ANTs 2.1.0)[53].

For each of the four BOLD runs per participant, the following preprocessing was performed. First, a reference volume and its skull-stripped version were generated using a custom methodology of fMRIPrep. The BOLD reference was then coregistered to the T1w reference using FSL flirt[54] with the boundary-based registration cost function. The coregistration was configured with nine degrees of freedom to account for remaining distortions in the BOLD reference. Head-motion parameters with respect to the BOLD reference (transformation matrices, and six corresponding rotation and translation parameters) were estimated before any spatiotemporal filtering using FSL mcflirt[54]. Automatic removal of motion artifacts using independent component analysis (ICA–AROMA) was performed on the preprocessed BOLD on MNI-space time series after removal of non-steady-state volumes and spatial smoothing with an isotropic, Gaussian kernel of 6 mm FWHM (full-width half-maximum). The BOLD time series were then resampled to the MNI152Nlin2009cAsym standard space.

The following 10 confounding variables generated by fMRIPrep were included as nuisance regressors: global signals extracted from within the cerebrospinal fluid, white matter, and whole-brain masks, framewise displacement, three translational motion parameters, and three rotational motion parameters.

**Cortical parcellation into brain regions.** We extracted neural responses across the whole brain for each video using the 200-parcel cortical parcellation scheme of Schaefer et al.[55] and 14 subcortical regions using the Harvard–Oxford subcortical atlas[56]. Together, this resulted in 214 regions that span the whole brain.

**Inter-subject correlations (ISCs).** We used the SciPy 1.5.3 library[57] in Python 3.7.0 to calculate ISCs. We extracted the mean time series in each of the 214 brain regions for each participant at each time point [i.e., at each repetition time (TR)]. Our analyses included 66 participants after the various exclusions, so there were 2,145 unique dyads. For each unique combination of dyad and video, we calculated the Pearson correlation between the mean time series of the neural response in each of the 214 brain regions. We then Fisher *z*-transformed the Pearson correlations and normalized the subsequent values (i.e., using *z*-scores) within each brain region.

**Relating neural similarity with information-sharing ratings**. As we described in the Results section, we wanted to test whether or not sharing likelihood is associated with neural similarity. To do this, we first binarized the sharing-likelihood ratings into a group with a high sharing likelihood and a group with a low sharing likelihood. The mean sharing-likelihood rating was 2.06 and the median was 2, so we classified sharing-likelihood ratings of 1 and 2 as "low likelihood" and sharing-likelihood ratings of at least 3 as "high likelihood". To relate this participant-level sharing-likelihood measure with the dyad-level neural-similarity measure, we transformed the participant-level sharing-likelihood measure for each video into a dyad-level measure for each video. We did this by creating a binary variable that indicated whether, for each video, both participants in a dyad had a high likelihood of sharing the video {high sharing, high sharing}, both participants had a low likelihood of sharing the video {low sharing, low sharing}, or one participant had a low likelihood of sharing the video and the other had a high likelihood of sharing it {low sharing, high sharing}. Of the 29,770 unique pairs of ratings, 3,485 were {high sharing, high sharing}, 14,963 were {low sharing, low sharing}, and 11,193 were {low sharing, high sharing}. We report analyses on a subset of the data using matched numbers of observations across the various sharing-likelihood levels in Supplementary Fig. 4.

To relate the dyad-level and video-level sharing-likelihood variables with neural similarity, we used the method in Chen et al.[28] and fit linear mixed-effects models with crossed random effects using LME4 and LMERTEST in R[58]. This approach allowed us to account for nonindependence in our data from repeated observations for each participant (i.e., because each participant is part of multiple dyads), each video (i.e., because each video was rated by multiple participants), and the interaction between each participant and each video (i.e., because each participant in a dyad rated each video). Following the method that was outlined in Chen et al.

(2017), we “doubled” the data (with redundancy) to allow fully-crossed random effects. In other words, we accounted for the symmetric nature of the ISC matrix and the fact that each participant contributes twice to each data point for each dyad (because $(i, j) = (j, i)$ for participants $i$ and $j$). We then manually corrected the degrees of freedom to $N - k$, where $N$ is the number of unique observations (in our case, $N = 29{,}770$) and $k$ is the number of fixed effects in the model, before performing statistical inference. All findings that we report in the present paper use the corrected number of degrees of freedom. For each of our 214 brain regions, we first fit a mixed-effects model, with ISCs in the corresponding brain region as the dependent variable and the dyad-level and video-level binarized sharing-likelihood variable as the independent variables, with random intercepts for each individual in a dyad (i.e., participant 1 and participant 2), each video, and the interaction between each individual and each video. We then conducted planned-contrasts using EMMEANS in R to identify the brain regions in which the ISCs were larger when participants indicated a higher likelihood to share a video than when they indicated a lower likelihood to share a video (i.e., $ISC_{\{high\ sharing,\ high\ sharing\}} > ISC_{\{low\ sharing,\ low\ sharing\}}$). In Supplementary Fig 1, we report results from the $ISC_{\{high\ sharing,\ high\ sharing\}} > ISC_{\{low\ sharing,\ high\ sharing\}}$ and $ISC_{\{low\ sharing,\ high\ sharing\}} > ISC_{\{low\ sharing,\ low\ sharing\}}$ contrasts. We converted all variables to z-scores to yield standardized coefficients (β) as outputs. We Holm–Bonferroni-corrected the $p$-values for multiple comparisons at $p < 0.05$.

**Study 2: Correlational behavioral study**

**Participants**. We recruited 100 participants who met our eligibility criteria, as outlined in our preregistration (see https://osf.io/qm4zw), on Amazon’s Mechanical Turk[59]. Participants were required to have an account on social media and to report that they sometimes share content (in this case, news stories) on social media. Specifically, to be eligible to participate, participants had to answer "yes" to both of the following questions: (1) "Do you currently have an account on any of the following social media platforms: Facebook, Twitter, Instagram?"; and (2) "Do you agree with the following statement? I sometimes share news stories on social media (for example, on Facebook, Twitter, and/or Instagram)." We determined our target sample size of this online convenience sample based on power calculations using pilot data, which suggested that we would have 95% power to detect a standardized effect size of $d = 0.13$, which was the smallest estimated effect size based on pilot data. The study was certified as exempt by the Institutional Review Board (IRB) of the University of California, Los Angeles. All participants saw an information sheet, in accordance with the procedures of UCLA's IRB.

**Procedure**. Participants completed an online survey that took 5–10 minutes and were compensated $0.85 after completing it. All participants saw the headlines and abstracts (i.e., short summaries) of five different news articles that were chosen uniformly at random from a sample of 29 news articles that we pretested in a pilot study to ensure that they (1) ranged in the extent to which their content would elicit similarity in interpretations across individuals and (2) were somewhat interesting, given that articles that are widely perceived to be uninteresting are unlikely to be shared (as a baseline) irrespective of how one believes others will interpret it. Hyperlinks to the stimuli are available at https://zenodo.org/records/13799055.

The order of the five news articles was assigned uniformly at random. Participants were asked their likelihood to share each article on social media with the question "How likely would you be to share this article on social media (e.g., on your Facebook timeline, Instagram, or Twitter)?" with the anchors "1 = extremely unlikely" and "5 = extremely likely". They were also asked the extent to which they believed that others in their social circles would have similar views as themselves about the article with the question "Consider the people with whom you are friends with on social media. How confident are you that they would all generally share your views on the content of the article?" with the anchors "0 = these people may or may not share my view" and "100 = I am confident that most of these people would share my view)". To counteract potential effects of seeing one type of question before the other, participants were assigned uniformly at random to see either all of the sharing questions first (and subsequently see all of the associated perceived-similarity questions) or all of the perceived-similarity questions first (and subsequently see all of the associated sharing questions). The order of the news articles was assigned uniformly at random for each set of questions. After answering all of the sharing and perceived-similarity questions, participants then rated how positive or negative they thought their friends on social media would find each article and how interesting they thought their friends on social media would find each article. For the first question, they were asked "To what extent do you think your friends on social media would view the content of each article in a positive or negative light?" with the anchors "0 = extremely negative", "50 = neutral", and "100 = extremely positive". For the second question, they were asked "To what extent do you think your friends on social media would find the content of each article interesting?" with the anchors "0 = extremely uninteresting", "50 = neither interesting nor uninteresting", and "100 = extremely interesting".

**Data analysis**. To test our main hypothesis that people are more likely to share content that they believe will be interpreted similarly by others in their social community, we fit a linear mixed-effects model using LME4 and LMERTEST in R[58]. This approach allowed us to account for nonindependence in our data from repeated observations for each participant (i.e., because each participant rated multiple news articles) and each news article (i.e., because each news article was rated by multiple participants). We fit a linear mixed-effects model with sharing likelihood as the dependent variable and perceived-similarity rating as the independent variable, with random intercepts for participant and news article. We also fit a linear mixed-effects model with sharing likelihood as the dependent variable, perceived-similarity rating as the independent variable, and participants' interest and valence ratings as control variables; we again used random intercepts for participant and news article. We converted all variables to z-scores to yield standardized coefficients (β) as outputs.

**Study 3: Behavioral experiment**

**Participants**. We recruited 300 participants on Prolific[60] who met the eligibility criteria, as outlined in our preregistration (see https://osf.io/7tvcb). Participants from this online convenience sample were required to be regular users of Facebook. (Specifically, they needed to use it at least once a month.) We determined our target sample size based on power calculations using pilot data, which suggested that we would have 85% power to detect a standardized effect size of $d = 0.25$, which was the smallest estimated effect size based on pilot data. The study was certified as exempt by UCLA's IRB, and all participants saw an information sheet, in accordance with the procedures of UCLA's IRB.

**Procedure.** Participants completed an online survey that took 5–10 minutes and were compensated $0.95 after completing it. Participants first filled out their demographic information, including their age, gender, race, socioeconomic status, sexual orientation, state of residence, political ideology, and political affiliation. They then provided their preferences on various topics, including their (unordered) top-three favorite movies of all time, their favorite and least-favorite sources of news, television shows that they found to be funny and not funny, and how they like to spend their free time. Participants were then presented with five news-article headlines and summaries and asked to select the one that most interested them. As in

Study 2, the five news articles were chosen to (1) range in the extent to which the content would elicit similarity in interpretations across individuals and (2) be somewhat interesting, given that articles that are widely perceived to be uninteresting are unlikely to be shared (as a baseline) irrespective of with whom one is considering sharing such articles. Hyperlinks to the stimuli are available at https://zenodo.org/records/13799122.

The participants were assigned uniformly at random into one of four conditions that manipulated how similar other members of a hypothetical Facebook group were to themselves: (1) similar social context, (2) dissimilar social context, (3) unclear social context, and (4) mixed social context. (See Supplementary Table 5 for the detailed instructions that the participants saw.) The participants then saw the news article that they had chosen earlier and were asked to indicate how likely they were to share that article with the Facebook group to which they were assigned. They were asked the question "How likely are you to share the following article with this Facebook group?" with the anchors "1 = extremely unlikely" and "5 = extremely likely". After providing their sharing-likelihood ratings, participants indicated how interesting they found the article to be and how often they typically share news articles on Facebook. For the first question, they were asked "How interesting is the following article to you?" (and they were again shown the article) with the anchors "1 = very uninteresting" and "5 = very interesting". For the second question, they were asked "How often do you share news articles on Facebook?" with the anchors "1 = less than once a year" and "5 = almost every day". We adopted our approach of experimentally assigning participants to different hypothetical Facebook groups from prior work[33].

**Data analysis**. To test our hypothesis that people are more likely to share content to others who they perceive as similar to themselves than to others who they perceive as dissimilar to themselves, we fit a linear regression model in R[61]. First, we fit a linear-regression model with sharing likelihood as the dependent variable and the experimental condition (i.e., social context) as the independent variable. We then conducted a planned-contrast analysis using EMMEANS in R[62] to test whether or not participants were more likely to share content with others who they perceived as similar than to others who they perceived as dissimilar (i.e., similar > dissimilar). We also examined all other possible contrasts in our framework (i.e., similar > mixed, similar > unclear, mixed > dissimilar, unclear > dissimilar, and mixed > unclear). We converted all variables to z-scores to yield standardized coefficients (β) as outputs. We FDR-corrected *p*-values for multiple comparisons at $p < 0.05$.

## Acknowledgements

We thank Meng Du, Elena Sternlicht, Kelly Xue, and the UCLA Center for Cognitive Neuroscience (particularly Jared Gilbert) for providing support with data collection. We thank Gang Chen for his statistical advice, and we thank the anonymous referees for their helpful comments. This work was supported by the National Science Foundation SBE Postdoctoral Research Fellowship [Grant No. 1911783] and the National Science Foundation [Grant No. SBE-1835239].

## Author Contributions

E.C.B., R.H., M.A.P., and C.P. designed the study and experiments. E.C.B., R.H., and K.L. collected the data. E.C.B. analyzed the data with support from C.P. E.C.B. and C.P. wrote the manuscript with feedback from all authors. C.B., C.P., and M.A.P. edited the manuscript.

## Competing Interests

The authors declare no competing interests.

## Data availability

The data that support the findings of this study are available from the corresponding author upon reasonable request.

## Code availability

The code used for the analyses is available from the corresponding author upon reasonable request.

Supplementary Information for **"Perceived community alignment increases information sharing"**

[1]Department of Psychology, University of Southern California, Los Angeles, CA, United States of America, [2]Department of Psychology, University of California, Los Angeles, Los Angeles, CA, United States of America, [3]Department of Mathematics, University of California, Los Angeles, Los Angeles, CA, United States of America, [4]Department of Sociology, University of California, Los Angeles, Los Angeles, CA, United States of America, [5]Sante Fe Institute, Santa Fe, NM, United States of America, [6]Brain Research Institute, University of California, Los Angeles, Los Angeles, CA, United States of America

* Corresponding authors:

Elisa C. Baek: elisa.baek@usc.edu

Carolyn Parkinson: cparkinson@ucla.edu

## Supplementary Table 1: Descriptions of stimuli

Supplementary Table 1. Descriptions of stimuli

| | Video | Content |
|---|---|---|
| 1 | An Astronaut's View of Earth | An astronaut discusses viewing Earth from space and, in particular, witnessing the effects of climate change from space. He then urges viewers to mobilize to address this issue. |
| 2 | All I Want | A sentimental music video depicting a social outcast with a facial deformity who is seeking companionship. |
| 3 | Scientific demonstration | An astronaut at the International Space Station demonstrates and explains what happens when one wrings out a waterlogged washcloth in space. |
| 4 | Food Inc. | An excerpt from a documentary discussing how the fast-food industry influences food production and farming practices in the United States. |
| 5 | We Can Be Heroes | An excerpt from a mockumentary-style series in which a man discusses why he nominated himself for the title of Australian of the Year. |
| 6 | Ban College Football | Journalists and athletes debate whether football should be banned as a college sport. |
| 7 | Soccer match | Highlights from a soccer match. |
| 8 | Ew! | A comedy skit in which grown men play teenage girls disgusted by the things around them. |
| 9 | Life's Too Short | An example of a 'cringe comedy' in which a dramatic actor is depicted unsuccessfully trying his hand at improvisational comedy. |
| 10 | America's Funniest Home Videos | A series of homemade video clips that depict examples of unintentional physical comedy arising from accidents. |
| 11 | Zima Blue | A philosophical, animated short set in a futuristic world. |
| 12 | Nathan For You | An episode from a 'docu-reality' comedy in which the host convinces people, who are not always in on the joke, to engage in a variety of strange behaviors. |
| 13 | College Party | An excerpt from a film depicting a party scene in which a bashful college student is pressured to drink alcohol. |
| 14 | Eighth Grade | Two excerpts from a film that depict a young teenager who video blogs about her mental-health issues and an awkward scene between two teenagers on a dinner date. |

Note: These videos were used in prior studies[1,2]; the descriptions of them in the present paper are the same as those in the prior studies.

## Supplementary methods 1 for analyses in Study 1

**Control variables.** As we noted in our descriptions of Study 1 in the Results section of the main manuscript, we controlled for the self-reported demographic variables in all of our models that related ISCs with sharing likelihood. These demographic variables consisted of participants' similarities in age, gender, and home country (which we define as the country in which an individual was living prior to enrolling at the university). To control for similarities in demographic variables, for each unique dyad (i.e., for each pair of individuals) in the fMRI session, we computed the absolute value of the difference between the ages of the two individuals in the dyad (i.e., $\text{age_difference} = |\text{age}_1 - \text{age}_2|$). We then transformed this difference score into a similarity score so that larger numbers indicate greater similarity (specifically, $\text{age_similarity} = 1 - (\text{age_difference}/\max(\text{age_difference}))$). To control for similarities in gender, we created an indicator variable in which 0 signifies different genders and 1 signifies the same gender. To control for similarities in home country, we used an indicator variable in which 0 signifies different home countries and 1 signifies the same home country. We then included these variables (i.e., similarities in age, gender, and home country) as control variables in our models that relate ISC and sharing likelihood.

**Permutation tests for sharing likelihood and video order.** As we noted in the main manuscript, all participants saw the videos in the same order. To address potential concerns that video order may affect sharing likelihood, we conducted permutation tests. Specifically, while holding sharing likelihood constant, we uniformly randomly shuffled the order of the videos 10,000 times. For each permutation of the data set, we calculated the Spearman rank correlation between the sharing likelihood and the labels that correspond to video order. This calculation generated an estimate of a null distribution of 10,000 Spearman correlation values that corresponds to what one would obtain by chance. We then computed a *p*-value by calculating the frequency with which the observed Spearman correlation between video order with sharing likelihood exceeded the Spearman correlation value in the null distribution. The observed Spearman correlation value of 0.054 did not differ from what one would expect based on chance, with a *p*-value of 1.000.

## Supplementary table for Study 1 results: Subcortical results

Supplementary Table 2. Results that relate ISCs with the binarized sharing variable: Subcortical results
Contrast: $ISC_{\{\text{high sharing, high sharing}\}} > ISC_{\{\text{low sharing, low sharing}\}}$

| Subcortical region | β | SE | *p* |
|---|---|---|---|
| Nucleus Accumbens (L) | –0.028 | 0.034 | 1.000 |
| Amygdala (L) | 0.035 | 0.051 | 1.000 |
| Caudate Nucleus (L) | –0.061 | 0.041 | 1.000 |
| Hippocampus (L) | 0.005 | 0.048 | 1.000 |
| Pallidum (L) | –0.056 | 0.030 | 1.000 |
| Putamen (L) | –0.084 | 0.035 | 0.551 |
| Thalamus (L) | 0.021 | 0.029 | 1.000 |
| Nucleus Accumbens (R) | 0.006 | 0.049 | 1.000 |
| Amygdala (R) | –0.025 | 0.043 | 1.000 |
| Caudate Nucleus (R) | –0.026 | 0.046 | 1.000 |
| Hippocampus (R) | 0.034 | 0.024 | 1.000 |
| Pallidum (R) | –0.025 | 0.024 | 1.000 |
| Putamen (R) | –0.105 | 0.040 | 0.044 |
| Thalamus (R) | 0.049 | 0.039 | 1.000 |

We Holm–Bonferroni-corrected all *p*-values due to multiple comparisons. The quantity β is the standardized regression coefficient, and SE is the standard error.

## Supplementary figure for Study 1 results: Results of exploratory contrasts

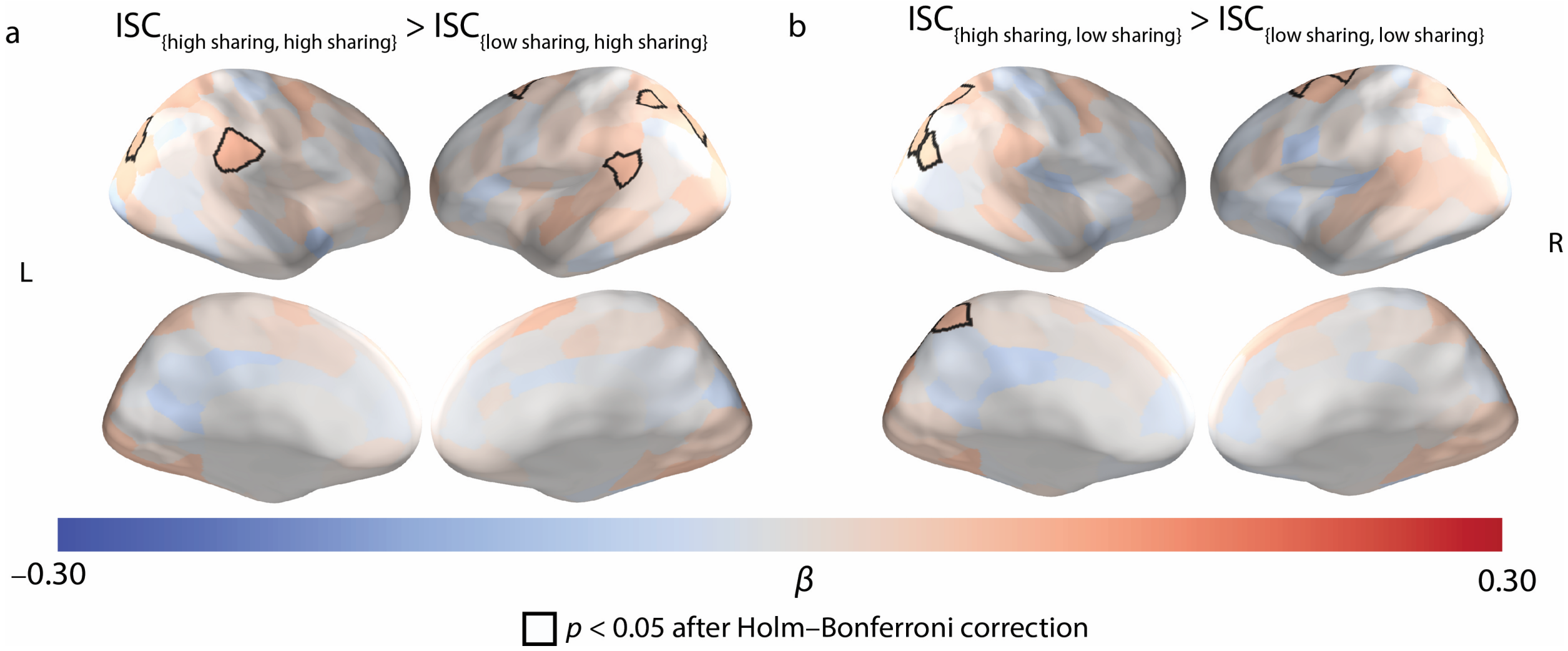

**Supplementary Fig 1. Additional exploratory contrasts that relate neural similarity with sharing likelihood.** **(a)** Similar to the results that we found for our primary contrast (i.e., $ISC_{\{\text{high sharing, high sharing}\}} > ISC_{\{\text{low sharing, low sharing}\}}$; see Fig. 1c in the main manuscript), we observed larger ISCs in the temporoparietal junction, the superior parietal cortex, and portions of the visual cortex when both participants were very likely to share content than when one participant was very likely to share content and the other participant was unlikely to share content (i.e., $ISC_{\{\text{high sharing, high sharing}\}} > ISC_{\{\text{low sharing, high sharing}\}}$). **(b)** We also obtained similar results for the $ISC_{\{\text{low sharing, high sharing}\}} > ISC_{\{\text{low sharing, low sharing}\}}$ contrast, with larger ISCs in portions of the visual cortex and superior parietal cortex when one participant was very likely to share content than when both participants were unlikely to share content. The quantity β is the standardized regression coefficient.

**Supplementary figure for Study 1 results: Results using a non-binarized version of the sharing-likelihood ratings**

In our primary analyses (which we reported in the main manuscript) of data from Study 1, we binarized our sharing-likelihood variable. The original sharing-likelihood variable was on a 1–5 Likert scale (with "1 = very unlikely" and "5 = very likely"). In our binarization, we classified ratings of 3 or more as a "high sharing likelihood" and ratings of 2 or less as a "low sharing likelihood". We also conducted analyses to test for associations between ISCs and a non-binarized version of the sharing-likelihood variable. To relate the participant-level sharing likelihood measure to the dyad-level neural-similarity measure, for each unique pair of participants, we first calculated a dyad-level variable that summarizes the overall likelihood of sharing each video by summing the sharing-likelihood ratings of both participants in a dyad. For example, if one member of a dyad rates their likelihood to share a video as "1" and the other member of the dyad rates their sharing likelihood as "4", then the dyad-level variable for sharing has the value 5. We then took an analogous approach to the one that we described in the Methods and Results sections for Study 1 results in the main manuscript. Specifically, for each of our 214 brain regions, we fit a linear mixed-effects model with crossed random effects with the ISC in the corresponding region as the dependent variable, the dyad-level non-binarized sharing-likelihood variable as the independent variable, and similarities in participants' age, gender, and country of origin as control variables. (See the Methods section of the main manuscript for more details on how we determined these control variables.) We also included random intercepts for each individual in a dyad, video, and interaction between each participant and each video. The models gave similar results (see Supplementary Fig. 2) as those that we reported in the main manuscript.

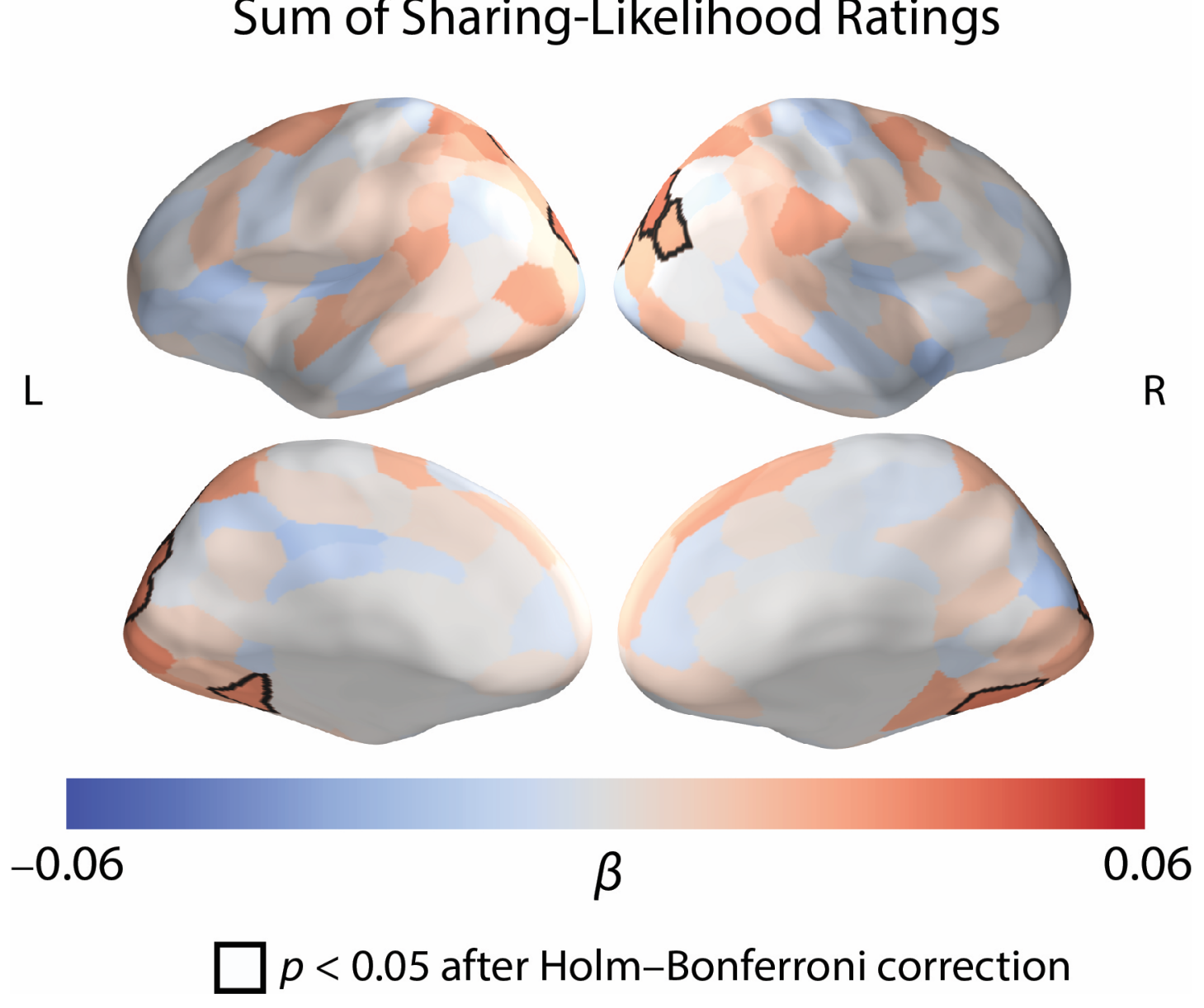

**Supplementary Fig 2. Relating neural similarity and sharing likelihood using a non-binarized variable to summarize sharing likelihood for members of each dyad. (a)** We observed similar patterns using a non-binarized version of the sharing-likelihood variable (which equals the sum of the sharing-likelihood ratings of the two participants in a dyad) as in the results that we obtained when we related a binarized version of the sharing-likelihood variable with ISCs (i.e., $ISC_{\{\text{high sharing, high sharing}\}} > ISC_{\{\text{low sharing, low sharing}\}}$; see Fig. 1c in the main manuscript). The quantity β is the standardized regression coefficient.

**Supplementary figure for Study 1 results: Alternative statistical-modeling approach**

To test the robustness of our results to alternative statistical-modeling approaches, we also conducted a modified version of the Mantel test of our primary analyses. First, we permuted the data 5000 times to create a null distribution of the data that accounts for the dependence structure of the data from repeating subjects and videos. Specifically, for each permutation, we first uniformly randomly shuffled the subject identifier while keeping the brain data constant. We then uniformly randomly shuffled the video identifier while keeping the brain data constant. We then ran the original mixed-effects model that we reported in the main manuscript. For each permutation, we added the maximum *t*-statistic across regions and contrasts to a null distribution. We then compared the actual *t*-statistics from the un-permuted data to those from the null distribution. We show the results of this procedure in Supplementary Fig. 3. The results from this permutation-based approach are similar to our findings with the method that we reported in the main manuscript, although fewer parcels emerged as significant using this approach.

We also attempted to model our data using the Bayesian multilevel-modeling approach of Chen et al.[3]. However, due to the complexity of our data structure and the large number of observations, it was not computationally feasible to deploy this Bayesian approach with the available resources.

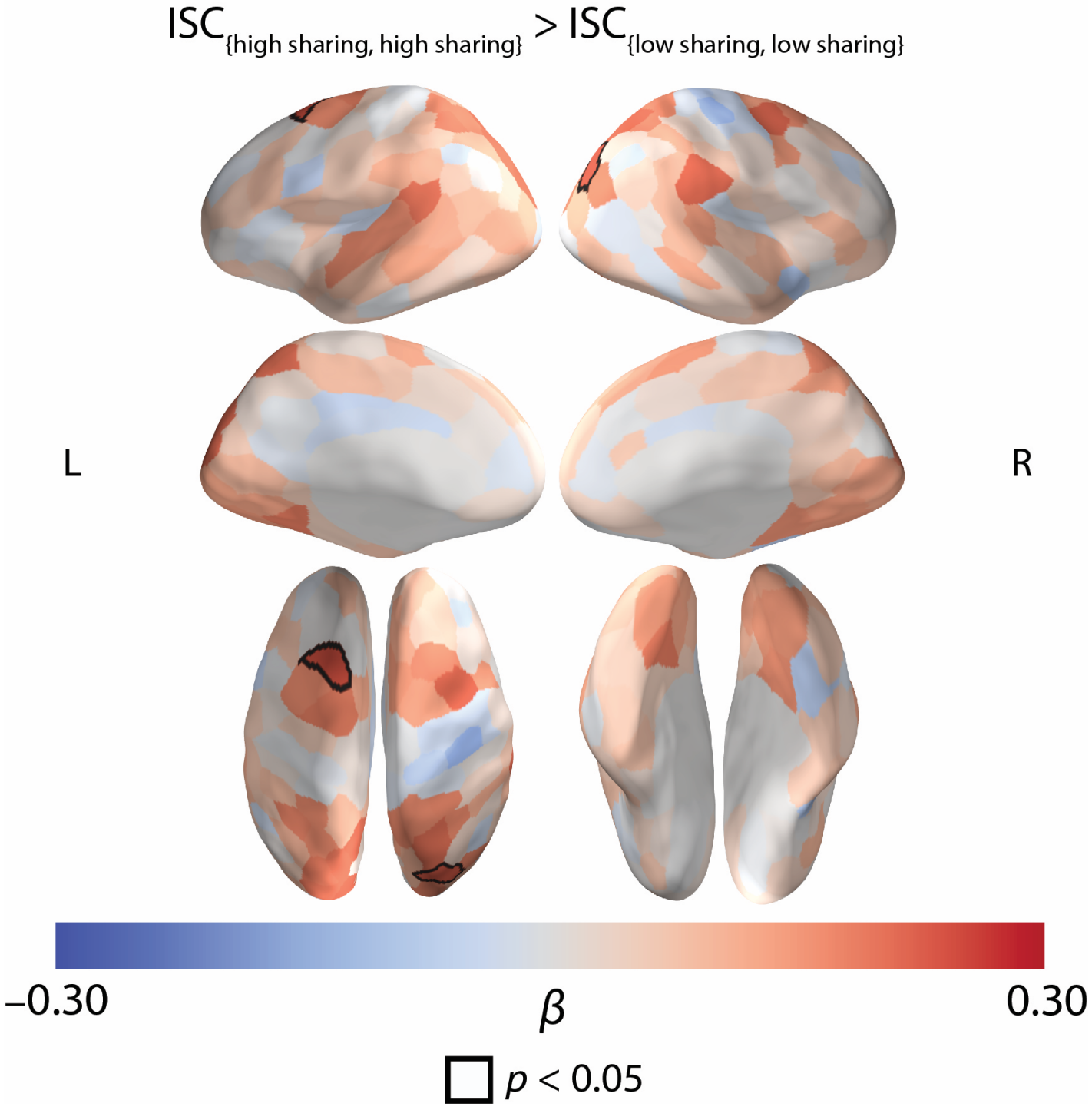

**Supplementary Fig 3. Relating neural similarity and sharing likelihood using a modified Mantel test with a maximum *t*-statistic approach.** Using this alternative statistical-modeling approach, we obtained a similar pattern of results to the ones that we reported in the main manuscript.

**Supplementary figure for Study 1 results: Results with matching numbers of observations**

Our dyad-level sharing-likelihood ratings had different numbers of observations for each of the three levels. Specifically, of the 29,770 unique pairs of ratings, 3,485 were {high sharing, high sharing} pairs, 14,963 were {low sharing, low sharing} pairs, and 11,193 were {low sharing, high sharing} pairs. We tested the robustness of the effects that we reported in the main manuscript to account for the different numbers of observations. To do this, we first undersampled the data uniformly at random from the {low sharing, low sharing} and {low sharing, high sharing} observations to match the number of {high sharing, high sharing} observations. We then fit our main model on this portion of the data set with matching observations across the different levels of the sharing variable. We repeated this process 1000 times. We then averaged across the 1000 estimates for the contrast of interest (namely, {high sharing, high sharing} > {low sharing, low sharing}). We show the results in Supplementary Fig. 4. As the figure indicates, the results of these analyses closely resemble the results that we reported in the main manuscript.

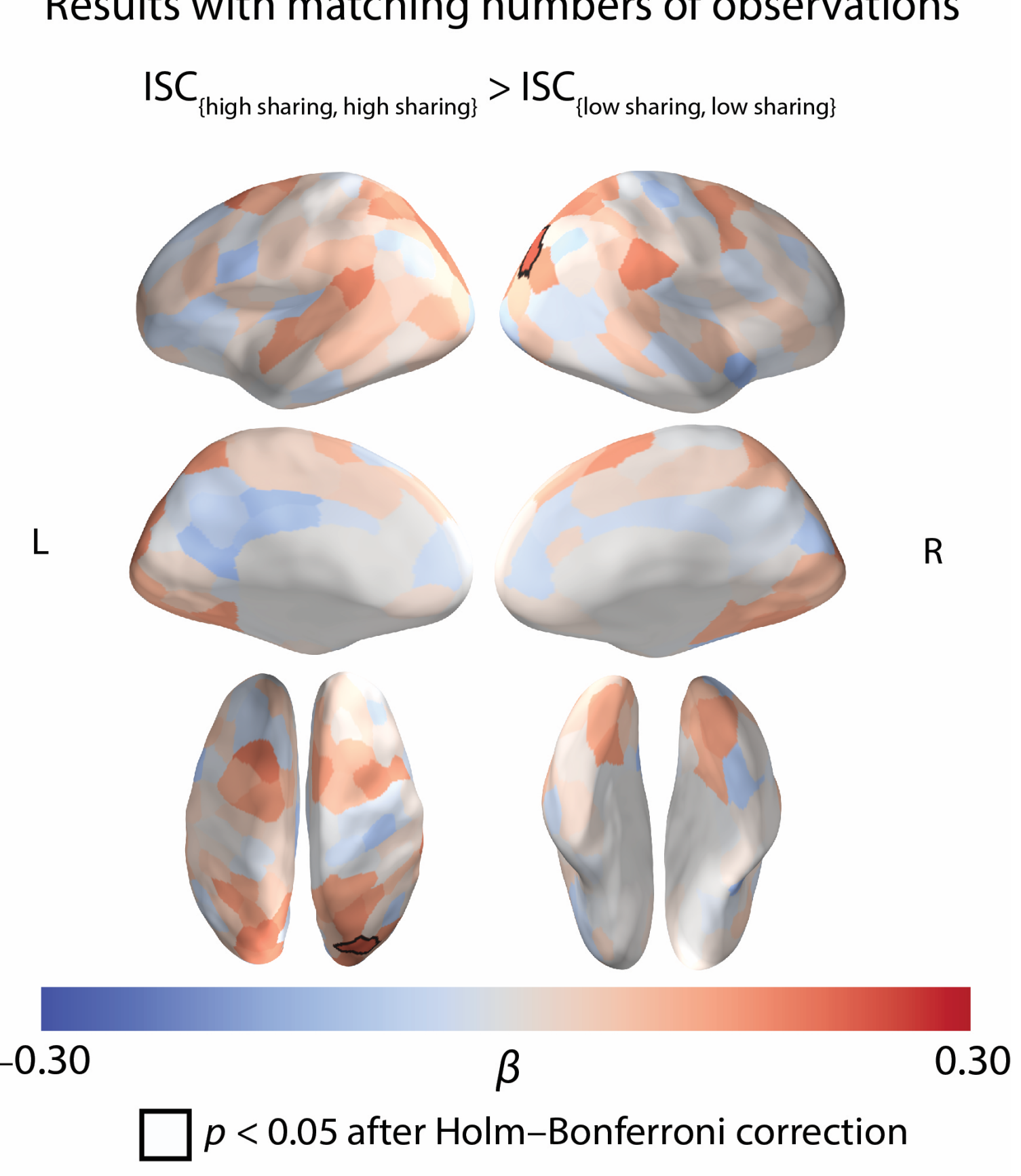

**Supplementary Fig 4. Relating neural similarity and sharing likelihood with matching numbers of observations.** The patterns of results that we obtained using subset data with matching observations of our dyad-level sharing-likelihood variable are similar to the ones that we reported in the main manuscript.

## Supplementary tables for Study 3 results: Results of all contrasts

Supplementary Table 3. Results of Study 3 for all contrasts for predicting sharing likelihood

| Contrast | β | SE | *p* |
|---|---|---|---|
| Similar > Dissimilar | 0.572 | 0.158 | 0.001 |
| Similar > Mixed | 0.289 | 0.158 | 0.092 |
| Similar > Unclear | 0.583 | 0.159 | 0.001 |
| Mixed > Dissimilar | 0.283 | 0.159 | 0.092 |
| Unclear > Dissimilar | –0.012 | 0.160 | 0.943 |
| Mixed > Unclear | 0.294 | 0.160 | 0.092 |

We have FDR-corrected all *p*-values due to multiple comparisons. The quantity β is the standardized regression coefficient, and SE is the standard error.

Supplementary Table 4. Results of Study 3 for all contrasts for predicting sharing likelihood when controlling for interest ratings and baseline sharing ratings

| Contrast | β | SE | *p* |
|---|---|---|---|
| Similar > Dissimilar | 0.854 | 0.192 | < 0.001 |
| Similar > Mixed | 0.534 | 0.192 | 0.012 |
| Similar > Unclear | 0.754 | 0.193 | < 0.001 |
| Mixed > Dissimilar | 0.320 | 0.193 | 0.147 |
| Unclear > Dissimilar | 0.010 | 0.195 | 0.608 |
| Mixed > Unclear | 0.220 | 0.195 | 0.312 |

We have FDR-corrected all *p*-values due to multiple comparisons. The quantity β is the standardized regression coefficient, and SE is the standard error.

## Supplementary table for Study 3 methods: Instructions for each experimental condition

Supplementary Table 5. Instructions for the different participant groups in Study 3

| Condition | Instructions |
|---|---|
| Similar | Now, imagine that you are invited to a group on Facebook by your colleagues.<br><br>When you join, you realize that the majority of people in this group are **similar to you** in your likes and dislikes about the things that you just provided your answers to. In other words, they share your sense of humor, favorite types of movies to watch, how they spend their free time, as well as in ideology and political leanings. |
| Dissimilar | Now, imagine that you are invited to a group on Facebook by your colleagues.<br>When you join, you realize that the majority of people in this group are **different from you** in your likes and dislikes about the things that you just provided your answers to. In other words, they **do not share** your sense of humor, favorite types of movies to watch, how they spend their free time, as well as in ideology and political leanings. |
| Mixed | Now, imagine that you are invited to a group on Facebook by your colleagues.<br>When you join, you realize that **some people in this group are similar to you and some people are different from you** in your likes and dislikes about the things that you just provided your answers to. In other words, **some people share** your sense of humor, favorite types of movies to watch, how they spend their free time, as well as in ideology and political leanings, but **other people do not**. |
| Unclear | Now, imagine that you are invited to a group on Facebook by your colleagues.<br>When you join, **you aren't sure** whether people in this group are similar to you in your likes and dislikes about the things that you just provided your answers to. In other words, you **aren't sure** whether they share your sense of humor, favorite types of movies to watch, how they spend their free time, as well as in ideology and political leanings. |